# Kondo destruction in a quantum paramagnet with magnetic frustration

Jiahao Zhang,[1,2] Hengcan Zhao,[1,2] Meng Lv,[1,2] Sile Hu,[1,2] Yosikazu Isikawa,[3] Yifeng Yang,[1,2,4] Qimiao Si,[5]

Frank Steglich,[1,6] and Peijie Sun[1,2]

[1] *Beijing National Laboratory for Condensed Matter Physics, Institute of Physics, Chinese Academy of Sciences, Beijing 100190, China*
[2] *University of Chinese Academy of Sciences, Beijing 100049, China*
[3] *Graduate School of Science and Engineering, University of Toyama, Toyama 930-8555, Japan*
[4] *Collaborative Innovation Center of Quantum Matter, Beijing, 100190, China*
[5] *Department of Physics and Astronomy, Rice University, Houston, TX 77005, USA*
[6] *Max Planck Institute for Chemical Physics of Solids, 01187 Dresden, Germany*

**Abstract**

We report results of isothermal magnetotransport and susceptibility measurements at elevated magnetic fields $B$ down to very low temperatures $T$ on high-quality single crystals of the frustrated Kondo-lattice system CePdAl. They reveal a $B^*(T)$ line within the paramagnetic part of the phase diagram. This line denotes a thermally broadened 'small' – to - 'large' Fermi surface crossover which substantially narrows upon cooling. At $B_0^* = B^*(T=0) = (4.6 \pm 0.1)$ T, this $B^*(T)$ line merges with two other crossover lines, viz. $T_p(B)$ below and $T_{FL}(B)$ above $B_0^*$. $T_p$ characterizes a frustration-dominated spin-liquid state, while $T_{FL}$ is the Fermi-liquid temperature associated with the lattice Kondo effect. Non-Fermi-liquid phenomena which are commonly observed near a 'Kondo destruction' quantum critical point cannot be resolved in CePdAl. Our observations reveal a rare case where Kondo coupling, frustration and quantum criticality are closely intertwined.

PACS:

Frustrated Kondo lattice systems have recently attracted much attention because of the emergence of exciting novel quantum phases. For example, in $Yb_2Pt_2Pb$, where frustration is caused by the quasi-two-dimensional Shastry-Sutherland lattice structure, a low-temperature phase emerges which most likely is a valence-bond solid [1]. Tetragonal $YbRh_2Si_2$ with inherent magnetic frustration shows, when substituted with several % of Ir (for Rh) [2] or Ge (for Si) [3], a non-Fermi liquid (NFL) phase in which the localized $4f$ – electron - derived magnetic moments are neither ordered nor Kondo screened. This identifies some critical quantum spin-liquid phase in the presence of a 'small' Fermi surface due to the ($s$, $p$, $d$) conduction electrons. Further examples for frustrated Kondo lattice systems are YbAgGe [4] and CeRhSn [5] with hexagonal ZrNiAl structure crystallizing in a distorted Kagome lattice with completely frustrated nearest-neighbor interactions.

In order to explore the interplay of the local moments' enhanced zero-point motion due to frustration and their Kondo coupling to the conduction electrons, a 'global' phase diagram (at $T = 0$) has been proposed which highlights fundamentally different paramagnetic and antiferromagnetic (AF) phases with either 'small' or 'large' Fermi surface (including also the $4f$ – electronic degrees of freedom) [6-8]. It contains a line of AF quantum critical points (QCPs), which intersects or even partially coincides with a 2[nd] order localization-delocalization quantum phase transition of the $4f$ electrons. The latter was predicted [9-11] to lead to an abrupt change of the Fermi-surface (or charge carrier concentration) which was indeed concluded from de-Haas-van-Alphen (dHvA) measurements on $CeRhIn_5$ under pressure [12-14] and isothermal magnetotransport [15, 16] as well as thermodynamic [17] measurements on $YbRh_2Si_2$. The finite-temperature signatures of this abrupt jump in the charge-carrier concentration at the QCP are thermally broadened changes in the afore-mentioned quantities at a crossover line $B^*(T)$ [18]. Interestingly, in *all* so-far investigated Kondo-lattice systems showing such a 'Kondo-destruction' instability, pronounced NFL phenomena in transport and thermodynamic quantities have been observed in the vicinity of this instability.

In this letter we address CePdAl which is isostructural to both YbAgGe and CeRhSn. The single crystals of CePdAl employed in this study were grown by the Czochralski technique in an argon atmosphere with induction heating. Details of the crystal growth were given in Ref. [19]. A single crystal from the same batch has been used for the thermodynamic measurements reported in Ref. [20]. Temperature and field-dependent longitudinal ($\rho_{xx}$) and transverse ($\rho_{xy}$) electrical resistivity as well as ac susceptibility ($\chi_{ac}$) were measured in a $^3$He-$^4$He dilution refrigerator down to $T$ = 0.08 K. In the present study, the magnetic field was always applied along the crystallographic $c$-direction. The excitation field in the $\chi_{ac}$ measurements was approximately 1 mT. The values of $\rho_{xx}$ and $\rho_{xy}$ were recorded by a Lakeshore 372 resistance bridge and those of $\chi_{ac}$ by a digital Lock-in amplifier (LI5460). The misalignment-induced uncertainty in transport measurements was minimized by measuring $\rho_{xx}$ and $\rho_{xy}$ on the same sample using a six-contact technique.

CePdAl has effective $S$ = 1/2 Ising-like, 4$f$-derived magnetic moments. As was previously shown for this compound, frustration gives rise to AF order involving only two thirds of the local moments at low fields and a sequence of three metamagnetic (MM) phase-transition lines at elevated fields [20-23]. The upper one at $B_{cp}$ ≈ 4.22 T, denoting an almost $T$-independent line of 1$^{st}$ order transitions below $T$ ≈ 0.8 K, separates AF order in 'phase $c$' from the paramagnetic regime.

Below, we will address the occurrence of a Fermi-surface crossover line $B^*(T)$ which terminates (as $T \rightarrow 0$) at a field $B_0^*$ = (4.6 ± 0.1) T (well separated from the AF order below $B_{cp}$). At this field, a crossover line $T_{FL}(B)$ terminates as well, which exists at $B$ > 4.6 T and below which a medium heavy Fermi-liquid (FL) is detected ['phase $f$', see Fig. 2]. Despite strong evidence for the existence of this 'Kondo-destruction' crossover at $B^*(T)$, no NFL phenomena are resolved in its vicinity down to the lowest measured temperature. Rather, a $T_p(B)$ line of crossovers into a low-$T$ '$p$-state' is found which, within the experimental uncertainty, also hits $B_0^*$ as $T \rightarrow 0$. Below this field, the '$p$-state' likely has a 'small' Fermi surface and local 4$f$ moments, hinting at some spin-liquid state. The phase diagram for CePdAl in the vicinity of its 'Kondo-destruction' instability (see discussion below) is strikingly different from the ones so far communicated for other materials [1-5].

Like in recent measurements of the specific heat ($C$) and magnetostriction ($\lambda$) [20, 23], a sequence of MM transitions at $B_{ab}(T)$, $B_{bc}(T)$ and $B_{cp}(T)$ is unambiguously revealed by low-$T$ isotherms of the ac magnetic susceptibility $\chi_{ac}(B)$ [Fig. 1a] as well as magnetoresistivity (MR) $\rho_{xx}(B)$ and its derivative d$\rho_{xx}(B)$/d$B$ [Fig. 1b]. A marked feature that has not been identified by previous thermodynamic probes is a 'shoulder' at $B^*(T)$ in $\chi_{ac}(B)$ and $\rho_{xx}(B)$ as well as a 'dip' in d$\rho_{xx}(B)$/d$B$ following, but well-separated from, the 1$^{st}$ order transition at $B$ = $B_{cp}$. The substantial reduction of MR at $B^*(T)$, which is also seen in the data of Ref. [24] (cf. Fig. 3b therein), is unaffected by the sharp drop at the 1$^{st}$ order transition, see Fig. 1b. As shown in this figure, we interpolate the d$\rho_{xx}(B)$/d$B$ data between the steep increasing line at $B_{cp}$ ≈ 4.22 T and the flat one above 4.8 T and subtract this interpolated (dashed black) line from the data measured in the same field window. This way, we obtain a minimum (red line), illustrating the derivative of the step-like drop in MR at $B$ = $B^*(T)$. It's full width at half maximum (FWHM) measures the crossover width. This finding is consistent with a field-induced increase of the charge carrier concentration at $B^*(T)$ due to the emergence of Kondo screening of the 4$f$ electrons [18]. We note that similar behavior at $B^*(T)$ has also been observed for YbRh$_2$Si$_2$ [25].

A contour plot mapping the value of d$\rho_{xx}(B)$/d$B$ measured at various temperatures (cf. Figs. 1b, S1 and Ref. [22]) is shown in Fig. 2. It includes all field-induced phases so far reported for CePdAl [20-24]. At temperatures below about 1 K, the whole phase space can be divided into three sectors: the one composed of different AF phases, '$a$' - '$c$' (left), the FL 'phase $f$' (right), and the intermediate paramagnetic region between the first-order phase-transition line at $B_{cp}$ and $T_{LF}(B)$, the $B^*(T)$ line being located within this intermediate region. As mentioned before, $B^*(T)$ was obtained from, e.g., the position of the minimum in $d\rho_{xx}/dB$ vs $B$, cf. Figs. 1b and S1. At $T \geq 0.8$ K, the three MM transitions converge into one $T_N(B)$ line, and the signature in $d\rho_{xx}(B)/dB$ at $B^*(T)$ evolves from a 'dip' into a broad minimum, see Fig. S1b. The dashed line $B_m(T)$, which (at a given temperature) marks the position of the maximum of the isothermal $\chi_{ac}(B)$ curve (Fig. S2), agrees well with the line of "maximum entropy" detected by Lucas et al. [23] who ascribe this to the putative long-range order phase-transition line in the absence of frustration.

The rich $B$ - $T$ phase diagram is further confirmed by isotherms of the Hall resistivity $\rho_{xy}(B)$ and its derivative $d\rho_{xy}(B)/dB$, cf. Fig. 1c. We assign the linear parts in $\rho_{xy}(B)$ at the lowest $T$, i.e., the horizontal parts in the derivative, in the windows below $B \approx 3.4$ T and above $\approx 5.5$ T, to the *normal* Hall coefficient $R_H(B)$, probing the concentration of hole-type majority charge carriers. There is a considerable reduction in $R_H(B)$, from 1.17 x $10^{-3}$ cm$^3$/C ('$a$' phase) to 0.65 x $10^{-3}$ cm$^3$/C ('$f$' phase), denoted by $R_{H,a}$ and $R_{H,f}$ respectively, cf. Fig. 1c. Based on an effective one-band model, this means a corresponding increase in the charge carrier concentration, on going from the low-field to the high-field window. Our observation that the delocalization of the Ce-derived 4$f$-electrons causes an increase of the effective *hole* concentration suggests a complicated renormalized band structure in the low-$T$ phase of CePdAl, as also reported for YbRh$_2$Si$_2$ [16, 26]. In the intermediate field range with multiple anomalies corresponding to the various phase boundaries as revealed by $\chi_{ac}(B)$ and $\rho_{xx}(B)$, the anomalous Hall effect is predominating.

The derivative of the Hall resistivity $d\rho_{xy}(B)/dB$ reveals a minimum at almost the same field, where shoulders are observed in both $\chi_{ac}(B)$ and $\rho_{xx}(B)$, cf. Figs. 1a and b. At the lowest temperatures $T = 0.08$ K (Fig. 1c) and 0.3 K (Fig. S3), this minimum at $B^*(T)$ is well-separated from the anomaly at $B_{cp}$. These observations suggest that the minimum of $d\rho_{xy}(B)/dB$ at $B^*(T)$, where the anomalous Hall effect is still present, is tracking the crossover in the underlying normal Hall coefficient, $R_H(B)$, between those low-field and high-field windows. Therefore, we attempted to use the position of this minimum and its FWHM to get empirical measures for the position of the $B^*(T)$ crossover and the crossover width, respectively. Indeed, both data sets agree well with the ones obtained from the resistivity (see Figs. 2 and 1d, right). Interestingly, at $T = 0.8$ K, Mochidzuki et al. [20] observe a broad hump in the isothermally determined specific-heat coefficient at a field (|| $c$) about 0.4 T higher than $B_{cp}$ (see their Fig. 3a), presumably identical with $B^*(T = 0.8$ K). Further experimental support for the $B^*(T)$ line is lent by the broad maxima in $\chi_{ac}(T)$ at $T^* = T(B^*)$ measured at constant fields, i.e., $B = 4.6$ T and 4.8 T (Fig. 4). A related $\chi_{ac}(T)$ maximum has been used in YbRh$_2$Si$_2$ [25] to identify the Fermi-surface crossover temperature $T^*(B)$ at a given field $B$.

In what follows, we shall provide evidence of this crossover line at $B^*(T)$ to extrapolate at zero temperature to a 2$^{nd}$ order quantum phase transition. In Fig. 1d (left panel), the $R_H$ values obtained from the plateaus in $d\rho_{xy}(B)/dB$ below 3.4 T and above 5.5 T (Figs. 1c and S3b) are presented as a function of temperature. There is no doubt that a considerable finite difference $\Delta R_H$ exists between these low- and high-field values when they are extrapolated to $T = 0$. In the right panel of this figure, we show the crossover widths obtained from the widths of the minima in both $d\rho_{xx}(B)/dB$ and $d\rho_{xy}(B)/B$ at $B^*$ as a function of temperature in a double-log presentation. These data are well fit by a power law dependence FWHM $\sim T^\varepsilon$ with $\varepsilon = (0.7 \pm 0.1)$ which suggests finite, discontinuous drops in both $\rho_{xx}(B)$ and $R_H(B)$ as $T \to 0$. This provides strong evidence for a sharp jump in the carrier concentration at $B_0^* \approx (4.6 \pm 0.1)$ T and is ascribed to an abrupt onset of Kondo screening, causing a delocalization of the 4$f$-electron states, on increasing the field through $B = B_0^*$. From the resistivity results displayed in Fig. 3, one infers that for $B \geq 4.8$ T the resistivity obeys a $T^2$ dependence, $\Delta\rho = (\rho - \rho_0) = AT^2$ ($\rho_0$: residual resistivity), below the crossover temperature $T_{FL}(B)$, indicated by the arrows in Fig. 3b (see also Fig. 2). The $A$-coefficient strongly increases when the field is lowered through $B_0^* \approx 4.6$ T (inset of Fig. 3b).

A Fermi-surface crossover line $B^*(T)$, separated from magnetic order but merging with the Fermi liquid crossover line $T_{FL}(B)$ at $T = 0$, agrees with current knowledge and has been observed, e.g., for Yb(Rh$_{0.94}$Ir$_{0.06}$)$_2$Si$_2$ [2]. An outstanding question that no prior system was suitable to address is the characterization of the paramagnetic phase in between $B_0^*$ and $B_{cp}$. With this in mind, we have measured the temperature dependence of the ac susceptibility $\chi_{ac}(T)$ in this field range. Surprisingly, $\chi'_{ac}(T)$ exhibits a pronounced increase upon cooling and tends to saturate at low temperature, see Fig. 4, consistent with a spin-liquid picture [29]. This property allows us to define a new temperature scale, $T_p(B)$, obtained from the inflection point of the increase in $\chi'_{ac}(T)$. The increase in the Pauli spin susceptibility as the temperature is lowered through $T_p$ suggests that it has the meaning of an effective Fermi energy for spin-carrying fermionic excitations in a metallic spin-liquid phase. We will refer to this low-temperature state as to the '$p$-state'.

The $T_p(B)$ line emerges at a multicritical point (4.2 T, 0.8 K), where the 1$^{st}$ order phase transition line (separating the '$c$-phase' from the '$p$-state') as well as the crossover line $B_m(T)$ terminate and the 2$^{nd}$ order dome-shaped phase boundary of the '$c$-phase' starts. $B_m(T)$ characterizes the change from the polarized paramagnetic phase to a state which exhibits short-range order of the frustration-induced type of AF order unique for CePdAl; therefore, a natural assumption is that also the '$p$-state' originates from magnetic frustration. This leads to the physical picture that as the temperature is lowered through $T_p$, the spins undergo a crossover from a regime of a partially Zeeman-split spin-liquid state (see the discussion below) to another one which is influenced by magnetic frustration. As seen in Fig. 3b, above $T_p(B)$ the resistivity follows a FL $T$-dependence, $\Delta\rho_{xx} = AT^2$ with an enhanced $A$-coefficient (inset of Fig. 3b), which resembles the case of pure YbRh$_2$Si$_2$ whose AF local-moment phase shows typical heavy FL behavior [25]. This has been ascribed to the dynamical Kondo screening which prevails at finite temperatures not only at the large but also at the small Fermi-surface side [18]. Below $T_p(B)$, the resistivity drops more rapidly, $\Delta\rho_{xx} \sim T^n$ with $n > 2$ (Figs. 3b and S4), suggesting additional freezing out of spin-flip scatterings due to frustration. Obviously, this new low-$T$ state in the quantum paramagnet CePdAl does not show any NFL $T$-dependence in the resistivity near $B_0^*$ and, in this sense, is very different from Yb(Rh$_{0.94}$Ir$_{0.06}$)$_2$Si$_2$ [2] as well as other systems studied so far [3, 5, 27, 28].

Remarkably, while the onset of the $p$-state crossover is still seen at 4.6 T (with a huge crossover width reaching all the way to $T = 0$), it is absent at 4.7 T. Therefore, within the experimental uncertainty, $T_p(B)$ merges with the $B^*(T)$ and $T_{FL}(B)$ crossover lines at $B_0^* = (4.6 \pm 0.1)$ T, i.e., at the Kondo-destroying instability (see also Fig. 2). This suggests that at $T = 0$, $B_0^*$ separates a paramagnetic phase with a small Fermi surface and an emerging energy scale $T_p(B)$ from a heavy fermion phase with a large Fermi surface and an emerging Fermi energy scale $T_{FL}(B)$.

To complete this picture, we note that both $B_0^*$ and $B_{cp}$ are large magnetic fields. Therefore, the Zeeman splitting in this field region considerably reduces the spin spectral weight associated with quantum fluctuations, as is illustrated by a broadened Schottky anomaly observed in the specific heat [20, 23], as well as a corresponding broad hump in the temperature derivative of the resistivity (Fig. S5) for fields $B > B_{cp}$. The Zeeman splitting has severe consequences on the physical properties of CePdAl: First, at $B > B_0^*$ and at temperatures above $T_{FL}$, the increase in the resistivity with $T$ becomes stronger than $T^2$ (see Figs. 3b and S4, for $B = 5$ T, 6 T and 7 T). This reflects the partial freezing of the spin-flip scatterings as temperature is reduced below the Zeeman-splitting scale. We attribute the lack of a linear temperature dependence in $\rho_{xx}(T)$ as well as other quantum-critical features near $B_0^*$ to this spin-flip freezing. Second, at low temperatures, when the field is increased towards $B_0^*$, the approach towards quantum criticality is accompanied by an increase of this partial spin-flip freezing. This is the reason for the overall trend of decreasing spin susceptibility with increasing $B$. Upon decreasing magnetic field, the $A$-coefficient of the $T^2$ term of the low-$T$ resistivity monotonously increases all the way up to the AF phase boundary at $B = B_{cp}$ (inset of Fig. 3b). Consistently, the overall 'background' of $\rho_{xx}(B)$ and $\chi_{ac}(B)$ underlying the various MM phase transition anomalies displays an apparent maximum at $B_{cp}$ (see Fig. 1a, 1b). The Sommerfeld coefficient of the electronic specific heat, too, is maximized at $B \approx B_{cp}$ [20,23]. These observations suggest that the '*c-p*' transition is only a *weak* 1st order transition. Very interestingly, as displayed in the inset of Fig. 3b, *A* vs *B* displays a steep variation near $B_0^*$, which indeed reflects the underlying Kondo destruction. Whether a smeared peak of *A* vs *B* exists at $B_0^*$ when the measurements are extended to even lower temperatures remains an intriguing open question.

To conclude, we have studied the field-induced paramagnetic part of the *T – B* phase diagram in the geometrically frustrated Kondo lattice system CePdAl. A well-defined crossover line $B^*(T)$ between a 'small' and a 'large' Fermi surface, terminating at a Kondo destroying instability, $B_0^* \approx 4.6$ T, has been identified. In addition, we have discovered a new crossover temperature scale $T_p(B)$, which characterizes the approach towards the small-Fermi-surface paramagnetic ground state at fields slightly below $B_0^*$. By connecting this crossover line with the crossover and phase transition lines at even lower fields, we are able to link the $T_p(B)$ line to the physics of magnetic frustration. The merging of this crossover line with $B^*(T)$ and the Fermi-liquid temperature line, defined at $B > B_0^*$, provides further support for a sudden jump of the Fermi surface at zero temperature across this critical field. We believe that the insight gained in our work will be relevant to a broader range of correlated metallic systems in which Kondo coupling, frustration and quantum criticality are closely intertwined.

**Acknowledgments**

We gratefully acknowledge valuable discussions with T. Xiang, J.L. Luo, L. Balents, H. von Löhneysen and P. Gegenwart. This work was supported by the Ministry of Science and Technology of China (Grant Nos: 2015CB921303, 2017YFA0303103), the National Science Foundation of China (Grant Nos: 11474332, 11774404), and the Chinese Academy of Sciences through the strategic priority research program (XDB07020200). The work at Rice was in part supported by the NSF grant No. DMR-1611392, the Robert A. Welch Foundation Grant No. C-1411, and the ARO Grant No. W911NF-14-1-0525.

**Figures and Captions:**

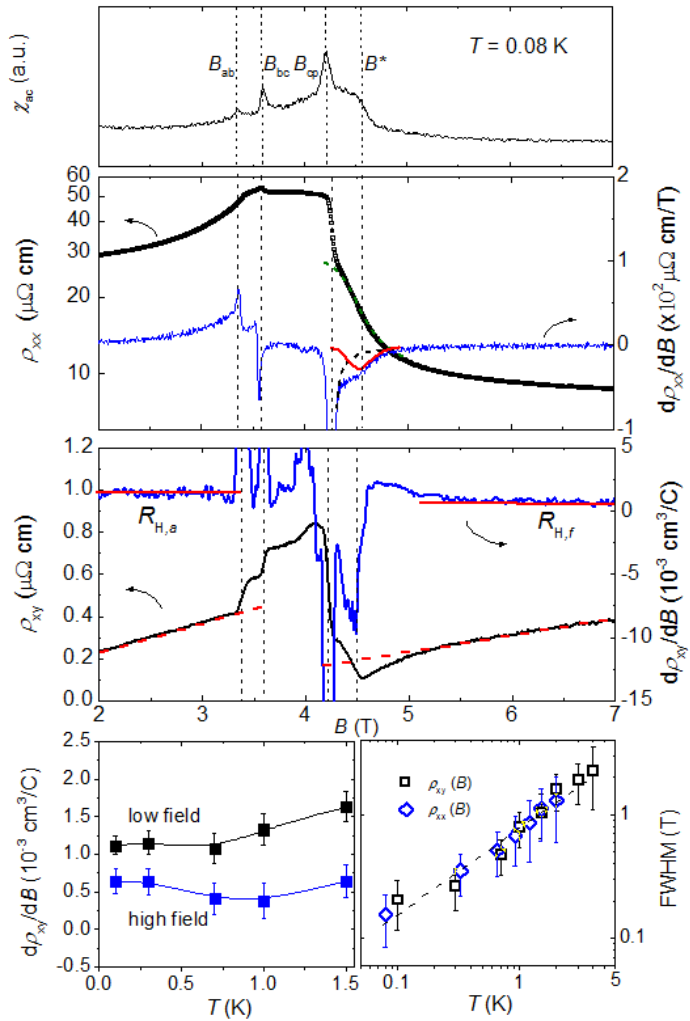

FIG. 1. **Characterization of the Fermi surface-crossover line $B^*(T)$** through low-$T$ isotherms of (**a**) ac susceptibility $\chi_{ac}(B)$, (**b**) magnetoresistivity $\rho_{xx}(B)$ and $d\rho_{xx}(B)/dB$ as well as (**c**) Hall resistivity $\rho_{xy}(B)$ and $d\rho_{xy}(B)/dB$. (**d** left) Values of Hall coefficient $R_H = d\rho_{xy}(B)/dB$ at lower (< 3.4 T) and higher (> 5.5 T) field, respectively [see (**c**)] as a function of $T$ and (**d** right) crossover width (FWHM) vs. $T$ derived from $d\rho_{xx}(B)/dB$ and $d\rho_{xy}(B)/dB$, see (**b**) and (**c**). As $T \to 0$, a finite difference $\Delta R_H$ ensues, but the crossover width vanishes. This is consistent with an abrupt Fermi-surface jump at $B = B_0^* = (4.6 \pm 0.1)$ T. The FWHM was obtained from both the deduced minimum in $d\rho_{xx}(B)/dB$ [red line in (**b**), see text] and the minimum seen in $d\rho_{xy}(B)/dB$ [see (**c**)]. In the intermediate field range 3.4 T < B < 4.3 T a sequence of three metamagnetic transitions at $B_{ab}$, $B_{bc}$ and $B_{cp}$ are indicated by distinct anomalies in all quantities exploited here, $d\rho_{xy}(B)/dB$ being dominated by the anomalous Hall effect in this field range.

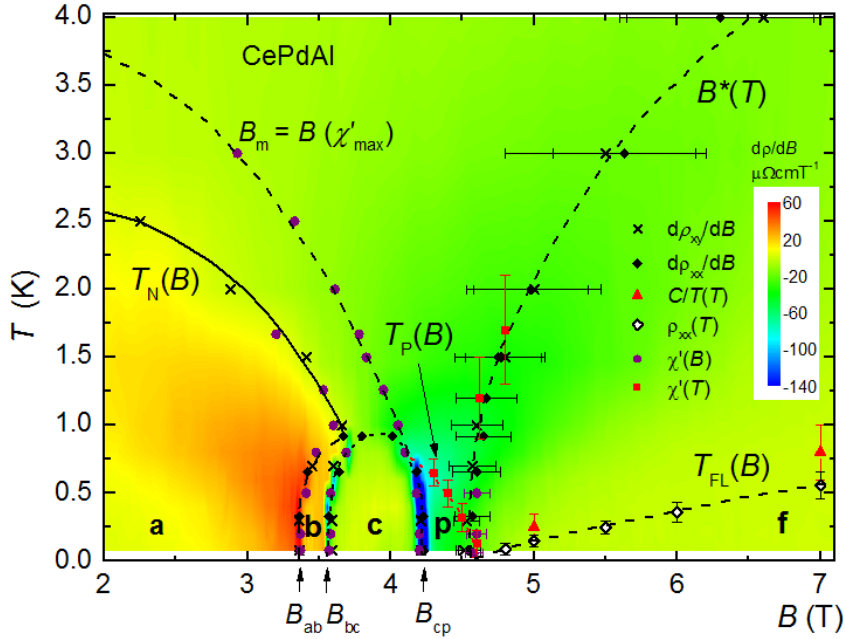

FIG. 2. **Temperature-field phase diagram of CePdAl.** Contour plot mapping the values of $d\rho_{xx}(B)/dB$ obtained for different temperatures, cf. Fig. 1b, 1S and Ref. [22]. Phase transitions/crossovers derived from various techniques are shown as well. This map is able to illustrate all the phases so far reported for CePdAl. The existence of a $B^*(T)$ line well separated from the magnetically ordered phases is based upon anomalies in $\chi_{ac}(B, T)$, $\rho_{xx}(B)$ and $\rho_{xy}(B)$. The emerging Fermi liquid (FL) phase ('$f$') at $B > B_0^* = (4.6 \pm 0.1)$ T can be identified from the onset of the $T^2$-dependent resistivity behavior (Fig. 3b), as well as from the inflection point in published results for the specific-heat coefficient $C(T)/T$ (red triangles) that connects a Schottky-type peak and a $T$-independent FL term [20, 23]. A low-temperature '$p$-state' exists below the $T_p(B)$ line of crossovers observed in $\chi_{ac}(T)$, see Fig. 4. It starts at a multicritical point (0.8 K, 4.2 T), where also the $B_{cp}(T)$ 1st order line and the $B_m(T)$ crossover line merge with the 2nd order line, that confines 'phase $c$' at elevated temperatures. The $B_m(T)$ line is defined by maxima in $\chi_{ac}(B)$ at $T \geq 1$ K (Fig. S2). It lies on top of the $T_S(B)$ line in Ref. [23], which characterizes maximum entropy and indicates the onset of frustration related AF short range order.

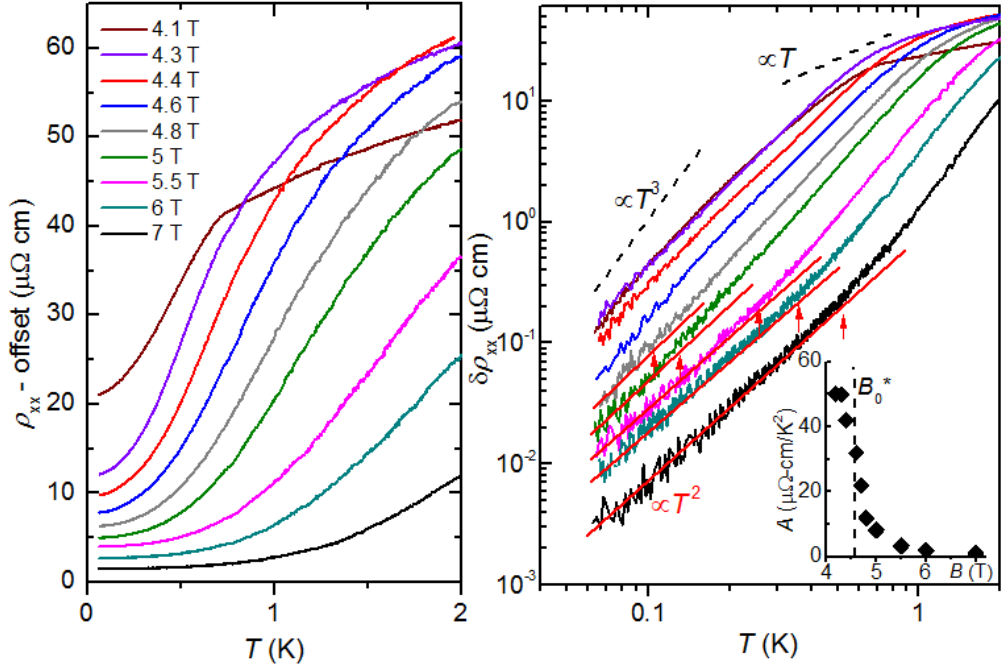

FIG. 3. **Temperature dependence of the low-$T$ resistivity for magnetic fields between 4 T and 7 T** in a (**a**) linear-linear and (**b**) double-logarithmic representation. Arrows in (**b**) mark the onset of a $T^2$ dependence, defining the $T_{FL}(B)$ line displayed in Fig. 2. Dashed lines illustrate ~$T$ and ~$T^3$ behavior. Inset of (**b**) shows the field dependence of $A$, the coefficient of $\Delta\rho_{xx} = A(B)T^2$ in the paramagnetic regime. It signifies an incipient divergence of $A(B)$ on the approach of the Kondo destruction instability at $B = B_0^* \approx 4.6$ T. Note that, for $B \leq B_0^*$, the value of $A$ was estimated by fitting $\Delta\rho_{xx}(T)$ in an appropriately higher temperature window (typically 0.2 – 0.5 K) where $\Delta\rho_{xx} \propto T^2$ holds, in order to avoid the low temperature regime ($T < 0.2$ K). Here, magnetic frustration dominates (Fig. S4), which inhibits the local moments from Kondo coupling to the conduction electrons.

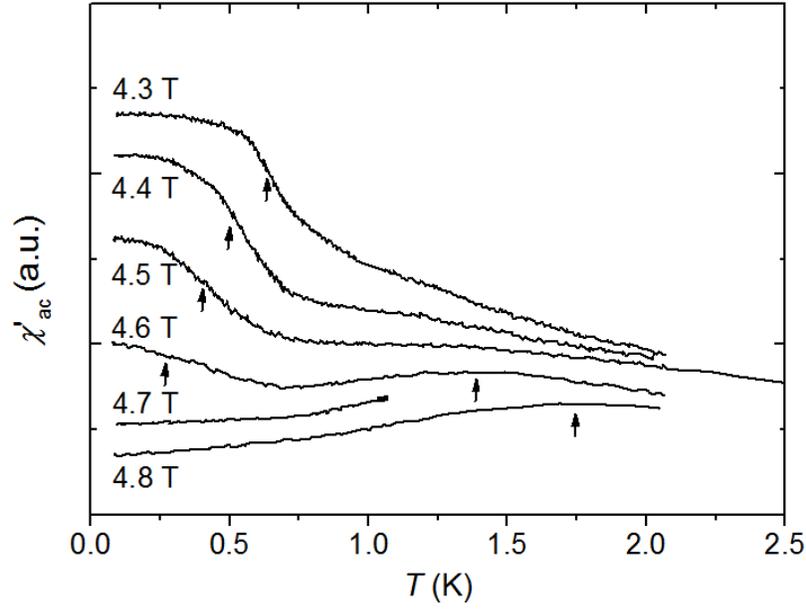

FIG. 4. **Low-$T$ ac susceptibility measured in various superimposed dc fields**. Shown is the real part of $\chi_{ac}(T)$. Arrows at the broad maxima of the $B$ = 4.6 T and 4.8 T curves denote the crossover at $T^*(B) = T(B^*)$, while arrows at lower temperatures, marking inflection points, signal the crossover into the '$p$-state' (see also Fig. 2), which exhibits a low-$T$ Pauli spin susceptibility. While it is absent at $B \geq 4.7$ T, the onset of the $p$-state crossover (with crossover width reaching to $T$ = 0) is observed at $B = B_0^* \approx 4.6$ T. No frequency dependence can be observed for the "shoulder structure" in the low-$T$ ac susceptibility [Fig. S6], ruling out spin-glass freezing.

# Supplementary Information.

1. **Isothermal resistivity.** Isotherms of electrical resistivity $\rho_{xx}(B)$ and its corresponding differential $d\rho_{xx}/dB$, which form the basis of the contour map in Fig. 2 of the main manuscript, are shown in Fig. S1. The three metamagnetic transitions at $B_{ab}$, $B_{bc}$, and $B_{cp}$ are clearly revealed by sharp anomalies in both $\rho_{xx}$ and $d\rho_{xx}/dB$ measured below $T \approx 0.8$ K. With increasing temperature, the emerging shoulder in $d\rho_{xx}/dB$ above these metamagnetic transitions gradually evolves into a broad minimum. This trend is revealed by the $B^*$ line shown in Fig. S1b. Fig. S1c illustrates how $B^*$ and the crossover width (FWHM) are determined from $d\rho_{xx}/dB$ vs $B$ curves, taking the measurement at $T = 0.66$ K as example, cf. Fig. 1b for similar analysis for $T = 0.08$ K. The dashed blue line is an extrapolation between the sharp minimum at $B_{cp}$ and the flat change above 5.5 T. The solid red line represents a difference between the extrapolated line and the measured values.

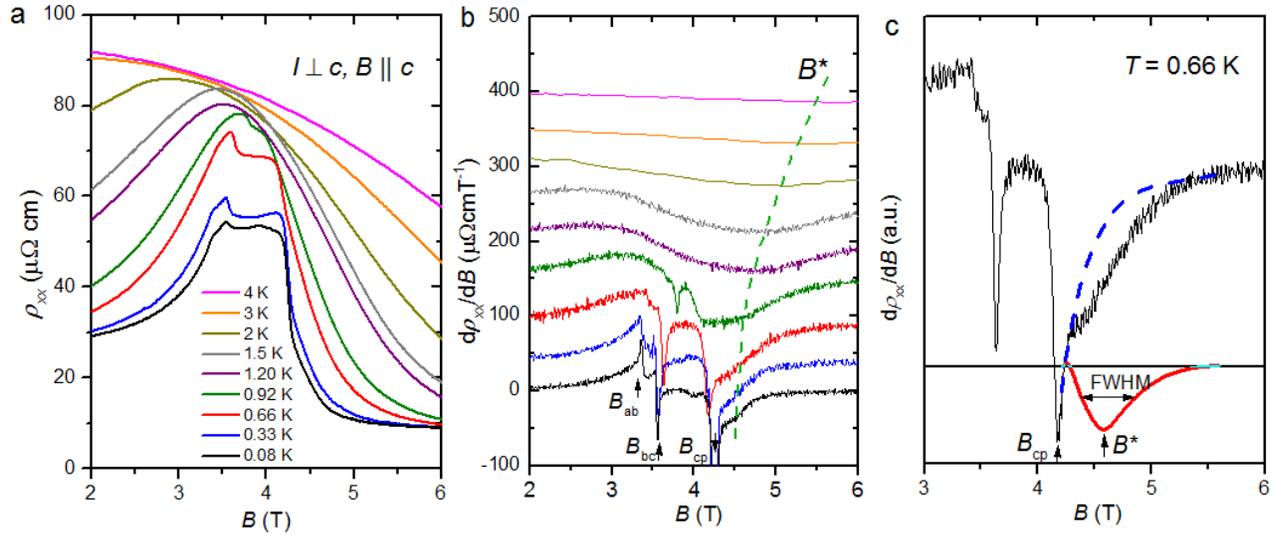

Fig. S1 Isothermal electrical resistivity $\rho_{xx}(B)$ (**a**) and its derivative $d\rho_{xx}/dB$ (**b**) measured at various temperatures. Panel (**c**) illustrates how $B^*$ and the crossover width are determined.

2. **Isothermal susceptibility.** The real part of the ac susceptibility (excitation field approximately 1 mT) isothermally measured as a function of a superimposed dc magnetic field at differing temperatures $0.08$ K $\leq T \leq 3$ K is shown in Fig. S2. Below 0.8 K, the three metamagnetic transitions at $B_{ab}$, $B_{bc}$ and $B_{cp}$ are clearly resolved. The shoulder at $B^*$ indicates the Fermi-surface crossover, see also Fig. 1a in the main manuscript. The positions of broad maxima in $\chi'_{ac}(B)$ for $T \geq 1.28$ K indicate a crossover at $B_m(T)$ associated with maximum entropy as discussed in Ref. [S1]. The kink below $B_m(T)$ for the curves taken at $T = 1.28$ K and 1.66 K denotes the onset of long-range antiferromagnetic (AF) order.

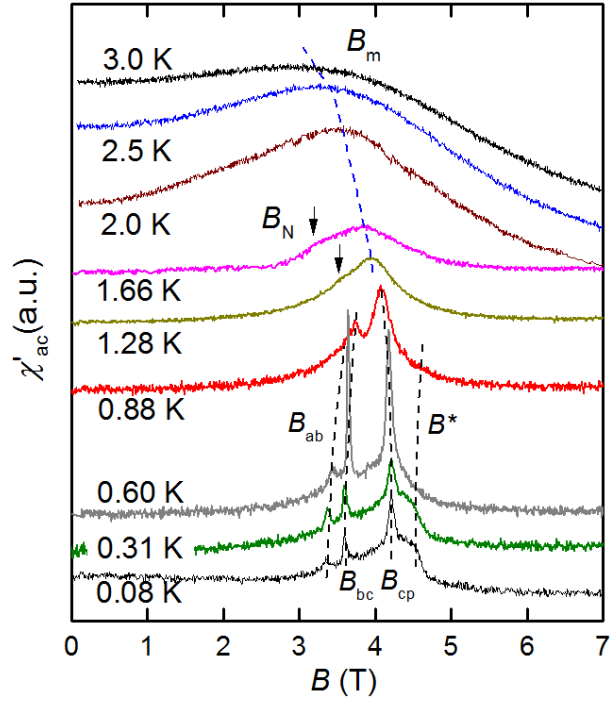

Fig. S2. Real part of ac susceptibility isothermally measured at various temperatures. Except for the metamagnetic transitions $B_{ab}$, $B_{bc}$, $B_{cp}$ and the crossover due to Kondo destruction at $B^*$, which have been shown in Fig. 1 of the main text, $B_m$ and $B_N$ are also indicated. The former marks the broad maximum in $\chi'_{ac}(B)$ measured at $T > 1$ K, the latter corresponds to the Neel critical field.

3. **Isothermal Hall resistivity.** Fig. S3 shows the isothermal Hall resistivity $\rho_{xx}(B)$ and its derivative $d\rho_{xy}(B)/dB$ measured between 0.08 K and 4 K. Except for the consequence of metamagnetic transitions $B_{ab}$, $B_{bc}$, $B_{cp}$ and the Neel critical field $B_N$, the isothermal $d\rho_{xy}(B)/dB$ curves provide a new identification of the Kondo destruction crossover at $B^*$. As has been discussed in the main manuscript, it reveals as an independent minimum in $d\rho_{xy}(B)/dB$ measured at $T$ = 0.08 K and 0.3 K, and evolves into a broad minimum at higher temperatures, cf. Fig. S3b.

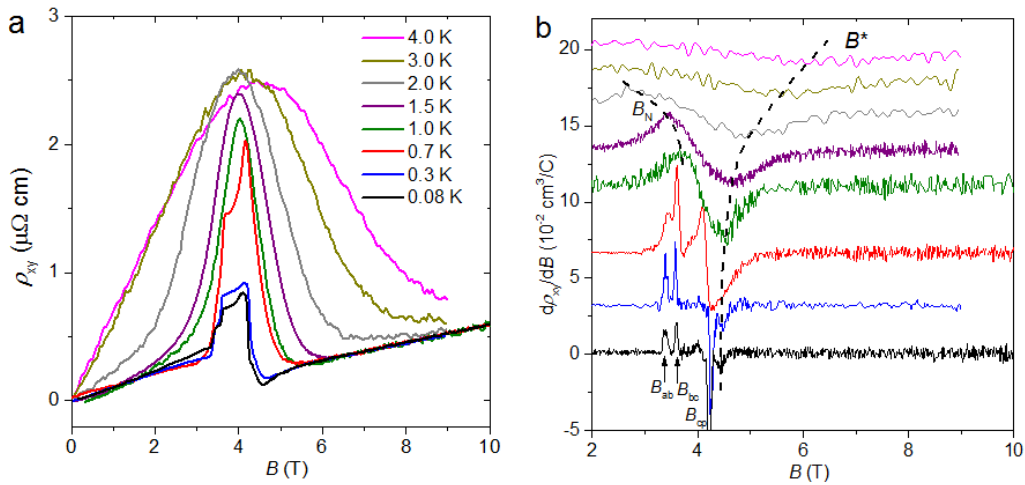

Fig. S3. Isotherms of Hall resistivity $\rho_{xy}(B)$ and the differential $d\rho_{xy}(B)/dB$ measured between $T$ = 0.08 and 4 K are shown in panel (**a**) and (**b**), respectively. Similar to the isothermal magnetic susceptibility (Fig. S2), in panel (**b**), the consequence of metamagnetic transitions and Neel critical field $B_N$ are indicated, as well as the crossover $B^*(T)$ line, which tracks the minimum in $d\rho_{xy}(B)/dB$.

4. **Temperature-field variation of resistivity exponent *n*.** Fitting the measured resistivity to $\delta\rho_{xx}(T) \propto AT^n$, the exponent *n* can be estimated following $n = \partial(\ln \delta\rho_{xx})/\partial(\ln T)$. Fig. S4 shows the contour map of *n*, where the phase transition/crossover lines are reproduced from Fig. 2. Refer to Fig. 3 in the main manuscript for the original $\delta\rho_{xx}(T)$ data. The blue coded regime above the $T_{FL}(B)$ line, where $n > 2$, arises from partial freezing of the spin-flip scatterings due to Zeeman splitting. The light-blue regime at the lowest temperatures ($T < 0.2$ K) in the '*p* state' reflects magnetic frustration.

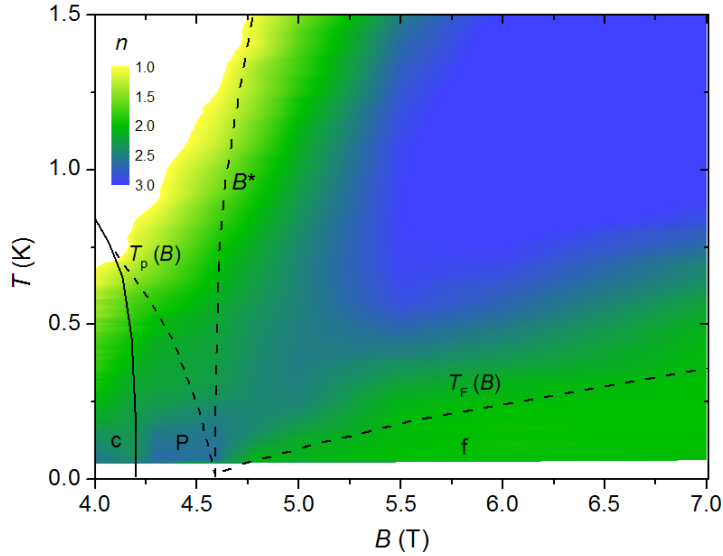

Fig. S4. Temperature-field variation of resistivity exponent *n* estimated from the $\Delta\rho_{xx}(T)$ curves shown in Fig. 3. The phase transition/crossover lines are reproduced from Fig. 2.

5. **Temperature derivative of resistivity.** Fig. S5 shows the temperature derivatives of resistivity, $d\rho_{xx}(T)/dT$, for selected fields $B > B_{cp}$. A broad maximum is observed for all the fields, which corresponds well to the Schottky anomaly in temperature-dependent specific heat induced by Zeeman splitting [S1]. This maximum in $d\rho_{xx}(T)/dT$ explains the $\Delta\rho_{xx}(T) = AT^n$ dependence with $n > 2$ at $T > T_{FL}$ ($B > B_0^*$). Below $B_0^*$ and $T_p(B)$, a similar power law dependence of $\Delta\rho_{xx}(T)$ is found. It is ascribed to dominating magnetic frustration which inhibits the local moments from Kondo coupling to the conduction electrons.

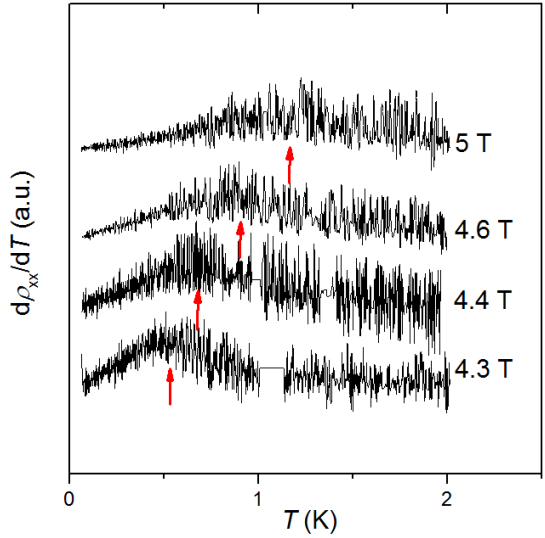

Fig. S5. Temperature derivative of resistivity, d$\rho_{xx}$/d$T(T)$, for selected fields above $B_{cp}$. The arrows mark the positions of maxima in the specific-heat coefficient that correspond to Schottky anomalies [S1].

6. **Frequency dependence of ac susceptibility.** Fig. S6 displays the results of $\chi_{ac}(T)$ measurements between 0.1 K and 1 K at a superimposed dc field of $B$ = 4.4 T. While the real part $\chi'_{ac}(T)$ does not dependent on the excitation frequency between 0.1 kHz and 2.5 kHz, the imaginary part $\chi''_{ac}(T)$ is completely temperature independent. These observations discard the possibility of the low-$T$ '$p$-phase' being due to spin-glass freezing [S2].

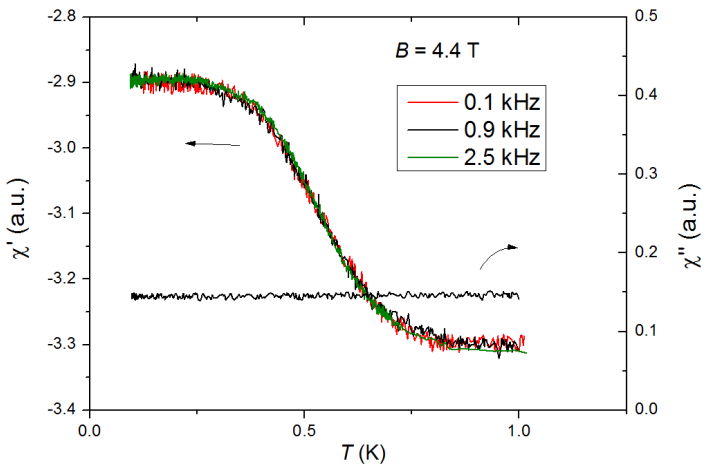

Fig. S6 Real (left axis) and imaginary (right axis) components of the ac magnetic susceptibility measured at different frequencies at an external superimposed dc field $B$= 4.4 T. Absence of frequency dependence and imaginary anomaly indicate the '$p$ state' is not a spin glass.

*[S1] S. Lucas et al., Phys. Rev. Lett. **118**, 107204 (2017).*

*[S2] J.A. Mydosh, Spin glasses: Redux: an updated experimental/materials survey, Rep. Prog. Phys. **78**, 052501 (2015).*